\documentclass[amsmath,amssymb,preprint]{revtex4}
\usepackage{graphicx}
\usepackage{slashbox}

\newcommand{\be}{\begin{equation}}
\newcommand{\ee}{\end{equation}}
\newcommand{\bea}{\begin{eqnarray}}
\newcommand{\eea}{\end{eqnarray}}

\begin{document}


\title{Parallel Algorithm for Calculation of the Exact Partition Function of a Lattice Polymer}

\author{Jae Hwan Lee}
\affiliation{School of Systems Biomedical Science and Department of Bioinformatics and Life Science, Soongsil University, Seoul 156-743, Korea}

\author{Seung-Yeon Kim}
\affiliation{School of Liberal Arts and Sciences, Chungju National University, Chungju 380-702, Korea}

\author{Julian Lee}
\email{jul@ssu.ac.kr}
\affiliation{School of Systems Biomedical Science and Department of Bioinformatics and Life Science, Soongsil University, Seoul 156-743, Korea}
\affiliation{Department of Pharmacuetical Chemistry and Graduate Group in Biophysics, University of California, San Francisco 94158, USA}

\date{\today}

\begin{abstract}
We develop a parallel algorithm that calculates the exact partition
function of a lattice polymer, by enumerating the number of
conformations for each energy level. An efficient parallelization of
the calculation is achieved by classifying the  conformations
according to the shape of the box spanned by a conformation, and enumerating only those in a given box at a time. The
calculation time for each box is reduced by preventing the conformations
related by symmetries from being generated more than once. The algorithm is
applied to study the collapse
transition of a lattice homopolymer on a square lattice,
by calculating the specific heat
for chain lengths up to 36.
\end{abstract}



\maketitle

\section{Introduction}
Polymers on discrete lattices serve as a simple toy model of a polymer~\cite{CD91,F49,F67,dG75,dG78,dG79}.
By introducing hydrophobic inter-monomer interaction, the lattice model can be used to study the collapse transition of a polymer~\cite{S75,B82,I82,KF84,BBE85,DS85,P86,S86,DS87,SS88,PCJS89,ML89,
CM93,GH95,BBG98,NKMR01,CL05,ZOZ08,CDC09,GV09}.
Various quantities such as radius of gyration, end-to-end distance, and specific heat have been calculated, both using Monte Carlo samplings~\cite{B82,BBE85,SS88,PCJS89,ML89,CM93,GH95,BBG98,NKMR01,ZOZ08,CDC09,GV09} and exact enumeration~\cite{I82,DS85,S86,CL05}.
Although the length of chain that can be studied using exact enumeration is much less than that of the Monte Carlo sampling methods, the exactness of the calculation enables one to use powerful extrapolation methods to study the behavior of the lattice polymers in the limit of infinite chain length.

 In the current study, we develop a method for calculating the exact partition function of a lattice polymer. The partition function is the most basic quantity from which all the important thermodynamic properties can be calculated. The number of states for each energy $E$, $\Omega(E)$, contains all the information needed for the calculation of the partition function $Z$, which is given by
\be
 Z(\beta) = \sum_E \Omega(E) e^{-\beta E},
 \ee
where $\beta \equiv 1/k_B T$ with the Boltzmann constant $k_B$ and temperature $T$.
 In particular, it can be used for studying the collapse phase transition of the lattice polymer by observing behaviors of various quantities such as  the specific heat,
\be
C(T,N) = \frac{\partial E}{\partial T} = \beta^2
\frac{\partial^2 \ln Z}{\partial \beta^2},
\ee
with increasing $N$.

 Since the number of
conformations is finite in the case of a lattice polymer, the
partition function can be made a polynomial in $e^{-\beta}$ by considering
models with integer values of $E$. To elaborate, we consider a polymer on regular lattice such as two-dimensional square or
three-dimensional cubic lattices. Then the Hamiltonian for a
heteropolymer is given by
\begin{equation}
    {\cal H} = - \sum_{i<j} \epsilon(a_i,a_j) \Delta ({\bf r}_i, {\bf r}_j),
\end{equation}
where
\be
	\Delta ({\bf r}_i, {\bf r}_j) = \left\{
              \begin{array}{ll}
                  1 & (|i-j| > 1 \quad {\rm and} \quad |{\bf r}_i-{\bf r}_j| =1)\\
                   0 & ({\rm otherwise}),
              \end{array}
       \right.
\ee
$a_i$ is the type of the $i$-th monomer, and $\epsilon(a,b)$ is the interaction energy between the monomers of type $a$ and $b$. By taking $\epsilon(a,b)$ to be integer multiples of a unit energy $\epsilon$, $\epsilon(a,b)=n(a,b)\epsilon$, the partition function is expressed as a polynomial:
\be
	Z = \sum_{\{\kappa(a,b)\}} \tilde \Omega(\{\kappa (a,b)\}) z^{\sum_{a,b}n(a,b) \kappa (a,b)}= \sum_K \Omega(K) z^K, \label{polpart}
\ee
where $z  \equiv \exp(\beta \epsilon) $, $\tilde \Omega(\{ \kappa (a,b)\})$ is the number of
polymer conformations with contact numbers $\{ \kappa (a,b)\}$, and
\be
\Omega(K) \equiv \sum_{\sum n(a,b) \kappa(a,b) = K} \tilde \Omega(\{ \kappa (a,b)\})
\ee
is the number of conformations with energy $E = -K \epsilon$.
The homopolymer composed of only one kind of monomers is a special case where $n(a,b)=1$ regardless of $a$, $b$, with $K$ now being the number of intra-chain contacts.

In this work, we develop an efficient parallel algorithm for calculating the exact partition function of a lattice polymer by enumerating $\Omega(K)$.
The parallelization is implemented by classifying the conformations by the shape of the box enclosing a conformation. Only the conformations corresponding to a given box is enumerated at a time by pruning partial conformations incompatible with the box, and the tasks of enumerating the conformations for the boxes are distributed among computational nodes.
Since no communications are required during the calculation, the computational speed
scales well with the number of CPUs.
The calculation time for each box is reduced by exploiting the symmetries of the system.
A conformation, its rotations by multiples of $90^\circ$, and their mirror images are considered equivalent
and prevented from being generated more than once.
For a generic conformation, the discrete rotations and reflections
form an 8-fold and 48-fold symmetries in two and three dimensions, respectively.
Exceptions are the cases of lower-dimensional conformations embedded in higher dimensional spaces.
In two dimensions, a straight chain is the one-dimensional conformation, invariant with respect to reflection perpendicular to the chain,
so the discrete rotations and reflections form a 4-fold symmetry.
Similarly, in three dimensions, only 6-fold and 24-fold symmetries exist for the
linear and planar conformations, since they are invariant under
transformation perpendicular to the underlying plane and straight line.

Therefore, the number of conformations with discrete rotations and reflections considered distinct in $D$ dimensions,
 $\Omega^{(D)}(K)$, can be easily obtained from the reduced number of conformations $\tilde \omega^{(D)}(K)$
 where symmetrically related conformations are counted only once
 and the lower-dimensional conformations are not counted:
\be
\begin{array}{l}
\Omega^{(1)} (K) = 2 \tilde \omega^{(1)} (K) = 2 \delta_{K,0}, \\
\Omega^{(2)} (K) = 8 \tilde \omega^{(2)}  (K)+ 4 \tilde \omega^{(1)}(K), \\
\Omega^{(3)} (K) = 48 \tilde \omega^{(3)}(K) + 24 \tilde \omega^{(2)}  (K)+ 6 \tilde \omega^{(1)}(K).
\end{array}
\label{total}
\ee
On the other hand, the reduced numbers of conformations $\omega^{(D)}(K)$ where only symmetries are eliminated, are expressed in terms of $\tilde \omega^{(D)}(K)$ as
\be
\begin{array}{l}
\omega^{(1)} (K) =  \tilde \omega^{(1)} (K) =  \delta_{K,0}, \\
\omega^{(2)} (K) = \tilde \omega^{(2)}  (K)+  \tilde \omega^{(1)}(K), \\
\omega^{(3)} (K) = \tilde \omega^{(3)} (K)  +  \tilde \omega^{(2)}  (K)+  \tilde \omega^{(1)}(K).
\end{array}
\label{reduced}
\ee
From Eqs.(\ref{total}) and (\ref{reduced}), we see that $\Omega^{(D)}(K)$ are expressed in terms of $\omega^{(D)} (K)$ as:
\be
\begin{array}{l}
\Omega^{(2)} (K) = 8 \omega^{(2)}  (K) - 4 \delta_{K,0} , \\
\Omega^{(3)} (K) = 48 \omega^{(3)} (K) - 24 \omega^{(2)}  (K) - 18  \delta_{K,0}.
\end{array}
\label{tot_red}
\ee
It is to be noted that the enumeration in two dimensions must be performed before enumerating those in three dimensions.

Although the current algorithm may be used for both homopolymer and heteropolymer in any dimension,
as a simple example of the application, we calculate the exact partition function for a homopolymer on a two-dimensional square lattice, for chain lengths up to 36. By analyzing the behavior of the specific heat, we could estimate the temperature of polymer collapse transition.

\section{The Method}

The crucial ingredients in the current method are
(1) classifying conformations according to the boxes they span, and enumerating only those for the given box at a time
by pruning partial conformations incompatible with the box, and
(2) preventing symmetrically related conformations from being generated more than once.
The arguments are almost the same for both two and three dimensions, but more detailed illustration will be given for two dimensions which is easier to visualize.

\subsection{The classification of the conformations according to the spanning boxes}

\begin{figure}
\includegraphics*[width=.9\textwidth]{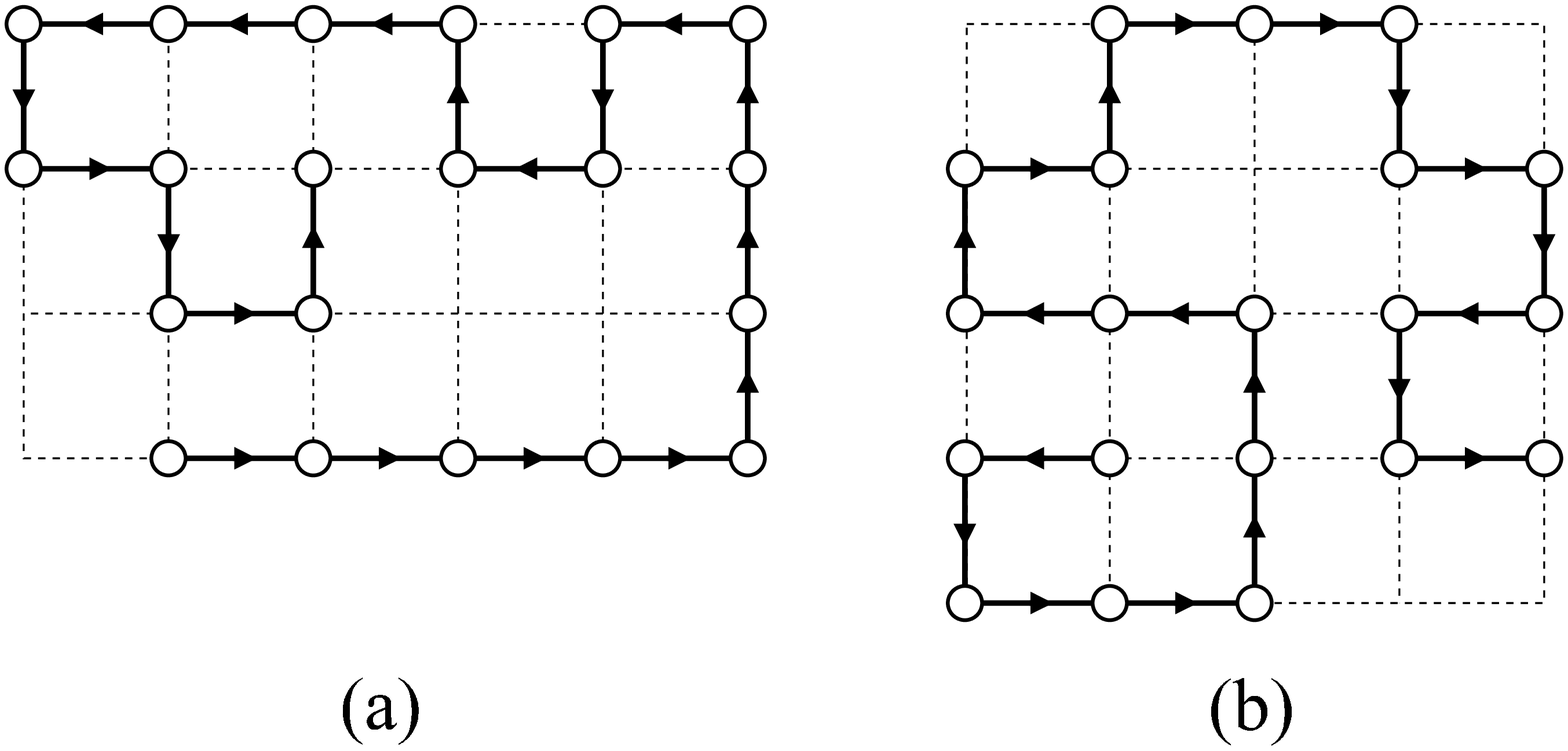}
\caption{Sample conformations for $N=20$ on
(a) $5 \times 3$ rectangle and (b) $4 \times 4$ rectangles.
Circles are monomers of a lattice polymer and the arrow indicates the direction of the chain.}
\label{sample}
\end{figure}

For each conformation, we construct a rectangular box enclosing it, whose sides are touched.
The conformations can then be classified according to the shapes of such boxes,
since the box is uniquely determined for each conformation.
The shape of a box can be described by the width $w$ and height $h$ in two dimensions,
and the depth $d$ is added for three dimensions.
We use the convention that these numbers are measured in the unit of the lattice spacing,
so the number of lattice sites spanned by the box is $(w+1)\cdot (h+1)\cdot (d+1)$.
For simplicity of the discussion, we keep $d$ regardless of the dimensions,
which is 0 for two dimensions.
As an illustration, conformations with box sizes $5 \times 3$ and $4 \times 4$ in two dimensions are depicted
in Figure~\ref{sample}. There is an intrinsic direction in the polymer chain, and a conformation for $N$ monomers can be considered as a self-avoiding walk of $N-1$ steps, starting from the 1st monomer.

\begin{figure}
\includegraphics*[width=1.0\textwidth]{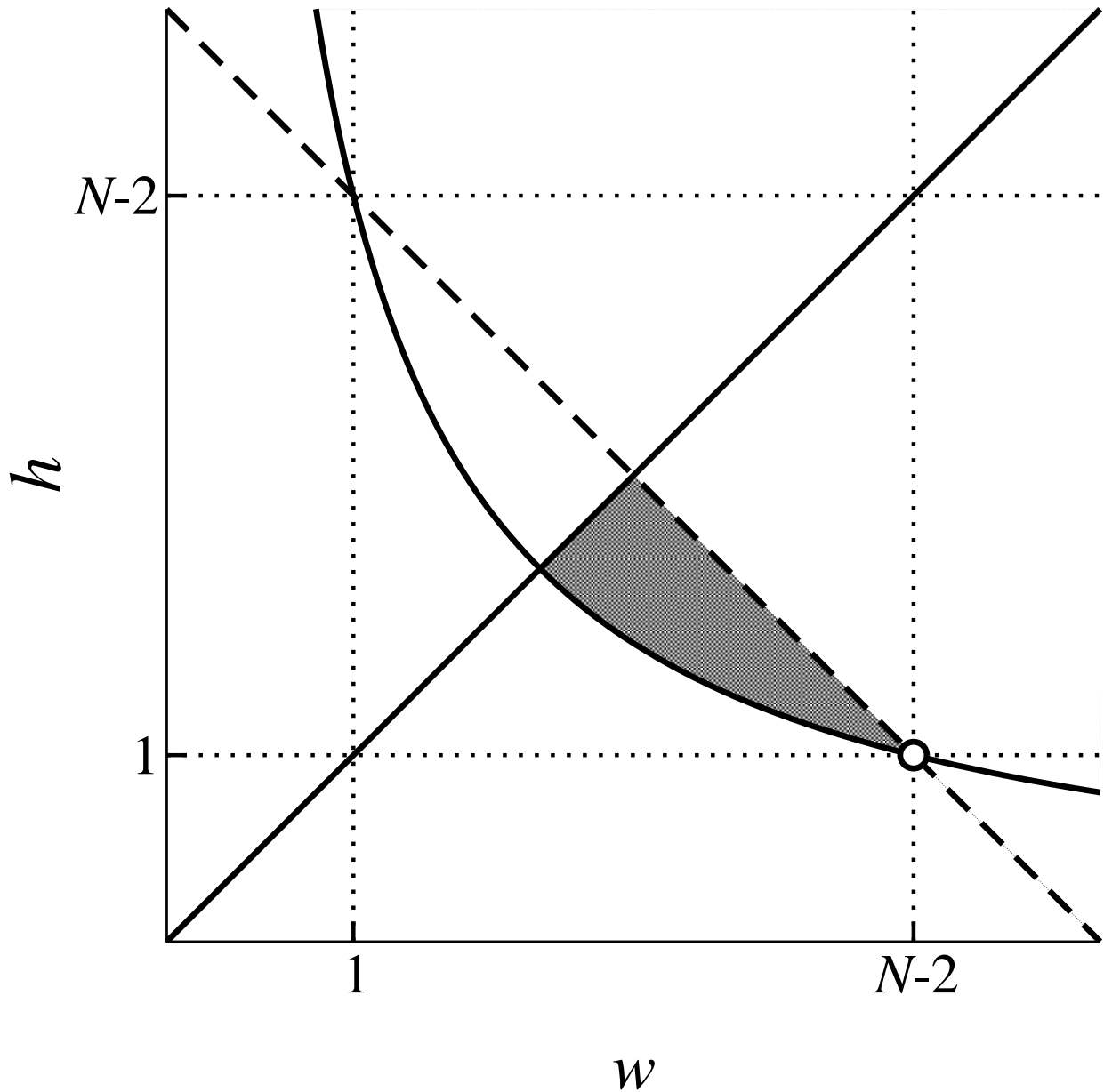}
\caption{The values of the box width $w$ and height $h$ for which the
explicit computations are performed in two dimensions, are the points with integer coordinates inside the shaded area. The dashed line and the intersection point denoted by the white circle is not included (see text). }
\label{range}
\end{figure}

Since the area or volume of a box should be large enough to
accommodate a conformation, $w$, $h$, and $d$ should satisfy the
lower bound $(w +1) \cdot (h + 1) \cdot (d+1) \ge N$  for a
chain with $N$ monomers. Also, since the perimeter of the box should
be small enough so that all sides touch the conformation, they should also
satisfy the upper bound $w + h + d \le N-1$. Due to discrete
symmetries, it is enough to consider only the boxes with $w \ge h
\ge d$.
Also, the conformations for boxes with $w$, $h$, and $d$ that saturate
the upper bound, $w+h+d=N-1$, do not have to be enumerated
explicitly, since they can be obtained from a simple analytic
formula. Considered as self-avoiding walks, these are the conformations where the steps are taken in a fixed direction along each axis, without turning back,  making no intra-chain contact at all.
 Since there are $D$ possible directions for each step in $D$ dimensions, the total
number of such conformations for all possible boxes is simply
$D^{N-1}/D!$ when rotational and reflectional symmetries are
eliminated.
Therefore, the values of $w$, $h$, and $d$ that have to be
included in the explicit enumeration are integers bounded by the
following inequalities:
\be
\begin{array}{c}
w \ge h \ge d \\
(w+1) \times (h+1) \times (d+1) \ge N \\
 w + h + d < N-1
\end{array}
\label{range_enum}
\ee
where again $d$ is set to 0 in two dimensions.
The region of $w$ and $h$ where explicit enumeration
is performed is shown in Figure~\ref{range} for the two-dimensional
homopolymer  along with the boundaries.

Only the conformations for a given box are enumerated at a time, and parallel computation is performed by distributing the boxes to the computational nodes. Since the number of boxes $N_{\rm box}$ is generally larger than the number of computational nodes $N_{\rm CPU}$, the simplest method of distribution would be to assign the same number of boxes to each node.
However, this method turns out not to be so efficient
since certain nodes finish jobs earlier and become free while others keep enumerating, due to the fact that the number of conformations and the computation time vary widely among the boxes.
 Therefore, the most efficient way of job
  distribution is to make enumeration for each box a separate task that can be taken up by any free node.
  Since a computational node that finishes the enumeration takes over a job in the
 queue, there is no idle time for any of the nodes.
The total number of boxes for each chain length must be precalculated in order
 for this efficient distribution, which can be performed in a
 practically negligible amount of time. Absolutely no communication is needed between the nodes during
 the enumeration, and it is only after all the computations are finished
 that the results from all the nodes are added
 to obtain $\Omega(K)$.
 The current algorithm can be used in a single
CPU as well, by requiring it to enumerate conformations for all
possible boxes.

\subsection{Self-avoiding walk and pruning}

The conformations are enumerated by generating self-avoiding
 walks. This is most
easily achieved by recursively calling a subroutine that
makes one step into a given direction~\cite{G93}. The number of sites $n$ occupied by the current
partial conformation is kept as a global variable, which is
equivalent to the number of lattice sites visited by the walk so
far. From the given position, the next step is made for each of the neighboring
lattice sites not occupied by a monomer. For each step of the walk,
the current lattice site is marked as occupied, and the contribution
of the current monomer is added to the number of contacts $K$. When  $n$
 reaches $N$, the contribution of the
final monomer is added to $K$ and $\omega(K)$ is incremented by one.

In order to ensure that the computation time is spent only for enumerating the conformations spanning a given box, any partial conformation incompatible with the current box is pruned out at an early stage. This is done by keeping the
record of whether each side of the box has already been touched by
the walk generated so far. If there is a boundary that has not been touched as yet, whose
distance from the current position is $l$, then the next step in the
opposite direction to this boundary is forbidden unless there are sufficient number of remaining monomers $N-n$.
Assuming the untouched boundary is at $x=0$ and the current position is at $x=l$ and the next step is in the positive $x$ direction, at least 3 monomers are needed to make a $U$-turn and get back to $x=l$,
and $l$ monomers are needed to reach the boundary $x=0$, leading to the inequality
\be
	l+3 \le N-n,
\label{prune_cond2}
\ee
which should hold for any step in the opposite direction from the untouched boundary, whose distance from the current position is $l$.
Similarly, the next step in the direction orthogonal to the
untouched boundary is forbidden unless
\be
	l+1 \le N-n.
\label{prune_cond2}
\ee
%
%

\subsection{Elimination of discrete symmetries}

\begin{figure}
\includegraphics*[width=.9\textwidth]{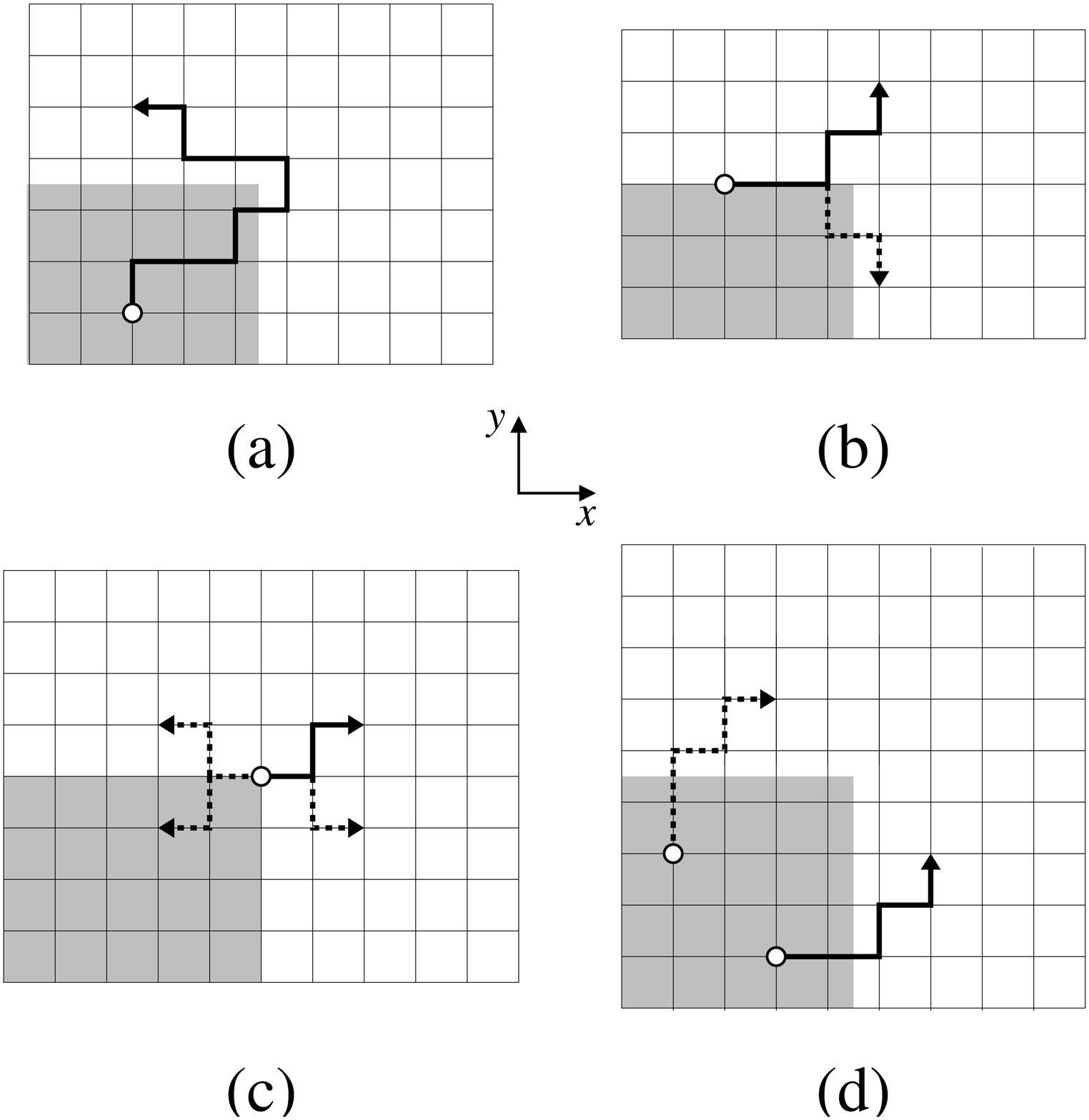}
\caption{(a) Because of the symmetries, only chains starting from the
sites in a shaded area at the lower left corner are considered.
(b) In the special case of chains starting at the boundary line
$y=\frac{h}{2}+1$, an additional constraint is imposed so that the
first vertical step is in the upper direction. A similar
constraint is imposed for a chain that starts at the boundaries
$x=\frac{w}{2}+1$ so that the first horizontal step is in the right
direction.
(c) For a chain starting at the center point with
$x=\frac{w}{2}+1$ and $y=\frac{h}{2}+1$ the two constraints for the
horizontal and vertical boundary are imposed simultaneously.
(d) For a square box, the additional reflectional symmetry with respect to
the line $y=x$ is eliminated by requiring that the first step is in
the horizontal direction.}
\label{rectangle}
\end{figure}

The speed of enumeration is increased by generating symmetrically
related conformations only once. Since the discrete symmetries of
rotation and reflection are 8-fold and 48-fold in two and three
dimensions respectively, the enumeration time is reduced by nearly
the same factor by calculating the reduced number of conformations
$\omega(K)$ instead of the full number $\Omega(K)$.
For a rectangular box where $w$, $h$, and $d$ are all different, $90^\circ$ rotational
symmetry is removed by considering only the box with $w > h > d$.
In order to remove the remaining 4-(8-)fold symmetry in two (three) dimensions,
we divide the rectangular box into 4 (8) equivalent quadrants (octants) containing each corner of the box,
and consider only the chain that starts inside one quadrant (octant), given by:
\be
\begin{array}{c}
1 \le x \le \frac{w}{2}+1 \\
1 \le y \le \frac{h}{2}+1 \\
1 \le z \le \frac{d}{2}+1,
\end{array}
\label{quadoct}
\ee
where $x$, $y$, and $z$ are integer coordinates for the lattice sites.
The quadrant described by Eq.~(\ref{quadoct})
 in the case of two-dimensions is shown as gray area in Figure~\ref{rectangle}(a).
Since symmetries are not completely eliminated for conformations starting at the boundaries,
where any of the conditions, Eq.~(\ref{quadoct}), is satisfied as an equality, additional constraints
are imposed, so that the first step along the axis corresponding to an equality is in the positive direction.
For example, when $w$ is an even number and if the starting position of the chain is at $x=\frac{w}{2}+1$,
we require that the first step along the $x$-axis should be positive.
Examples in two dimensions are shown in Figures~\ref{rectangle}(b) and (c).

When any two of the three numbers $w$, $h$, and $d$ are equal, for example $w=h$, there remains a reflection symmetry
with respect to plane $x=y$,  which is eliminated by imposing an
additional constraint that among the steps parallel to the $x$-$y$ plane, the first one must be along the $x$ direction. Similarly,
 when $h=d$ we break the symmetry by requiring that among the steps parallel to the $y$-$z$ plane, the first step must be along the $y$ axis.  Again the two-dimensional example is given in the Figure ~\ref{rectangle}(d).

\section{Result}

The method developed in the current work can be applied to either heteropolymer or homopolymer in any dimension,
but as a simple example of the application,
we study the homopolymer in the two-dimensional square lattice.
Since a monomer cannot make a contact with itself, as well as the nearest and next-nearest neighbors along the chain, the upper limit of the $K$ summation in Eq.~(\ref{polpart}), denoted as $K_{\rm max}$, satisfies the upper bound
\be
	K_{\rm max} \le \frac{N(N-5) n_{\rm max}}{2},
\ee
where $n_{\rm max}$ is the maximum value of $n(a,b)$.  In the absence of an additional information, we may take this value  as the size of the array to store the values of $\omega(K)$, but for the  special case of the homopolymer in two dimensions, we have the exact formula for  $K_{\rm max}$~\cite{CD89}:
\begin{equation}
    K_{\max} = \left\{
    \begin{array}{ll}
      N-2m & {\rm for} ~~ m^2 < N \le m(m+1),\\
      N-2m-1 ~~~ & {\rm for} ~~ m(m+1) < N \le (m+1)^2 ,
    \end{array} \right.
\end{equation}
where $m$ is a positive integer.

\begin{figure}
\includegraphics*[width=.9\textwidth]{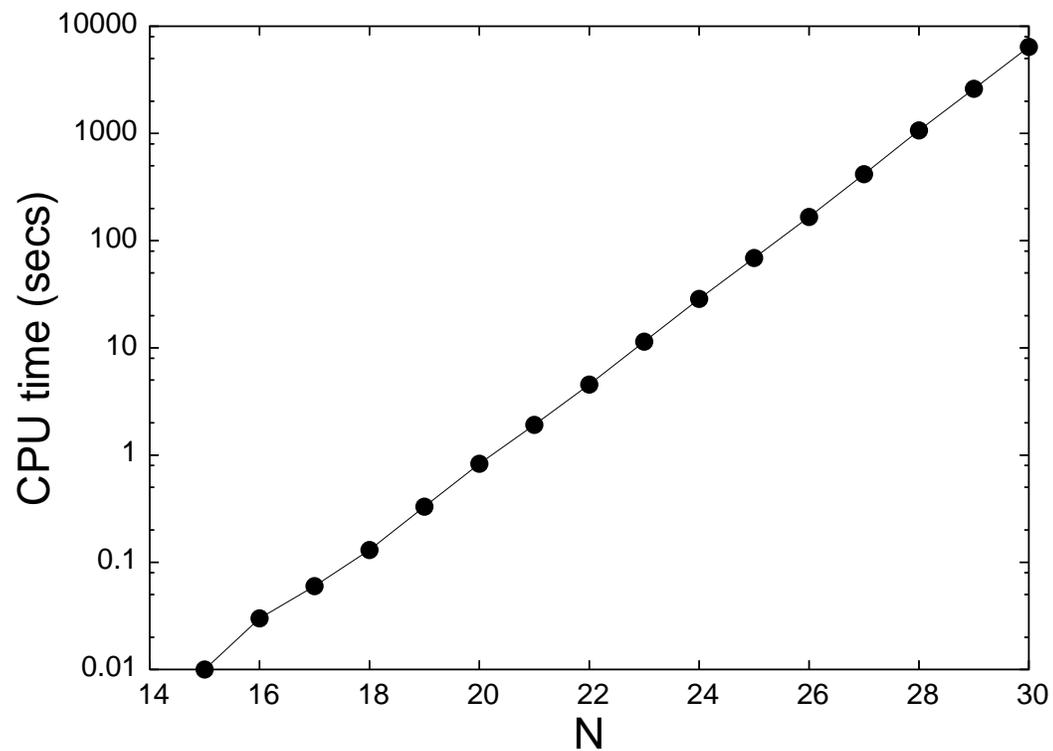}
\caption{The CPU time in log scales, plotted as a function of the chain length $N$.}
\label{time}
\end{figure}

We calculated the number of states $\omega(K)$ for $N \le 36$. The same quantities for $ N \le 28$ have been calculated in earlier works~\cite{CD89,CD91,L04}. Our calculation reproduce these results in the appropriate ranges.
The new results for $29 \le N \le 36$ are presented in Table~\ref{29to32} and \ref{33to36}.
The CPU time using Intel Xeon CPUs (2.8GHz) is plotted in logarithmic scale as the function of $N$ in Figure~\ref{time}, for $15 \le N \le 30$ where time data are available. As expected, the CPU time grows exponentially as $N$ increases.
\be
	t \simeq A \lambda^N
\label{timefunc}
\ee
where values of $A=1.64(2) \times 10^{-8}$ and $\lambda=2.43(2)$ are obtained
by taking the log of Eq.~(\ref{timefunc}) and performing  the least square fit.
The total number of conformations as well as the computational times of many enumeration algorithms are known to grow exponentially as $\mu^N$ where $\mu=2.638$ is the connective constant for the self-avoiding walks~\cite{JC01},
so the computational time of the current algorithm grows at somewhat slower rate than this. Although algorithms for enumerating the {\it total} number of conformations  have been developed~\cite{EN80,J04} whose computation time grows at rates slower than the current one, it must be noted that the current method  not only calculates the total number but also the number of conformations for each energy value, leading the calculation of the exact partition function at arbitrary temperature.

\begin{figure}
\includegraphics*[width=.9\textwidth]{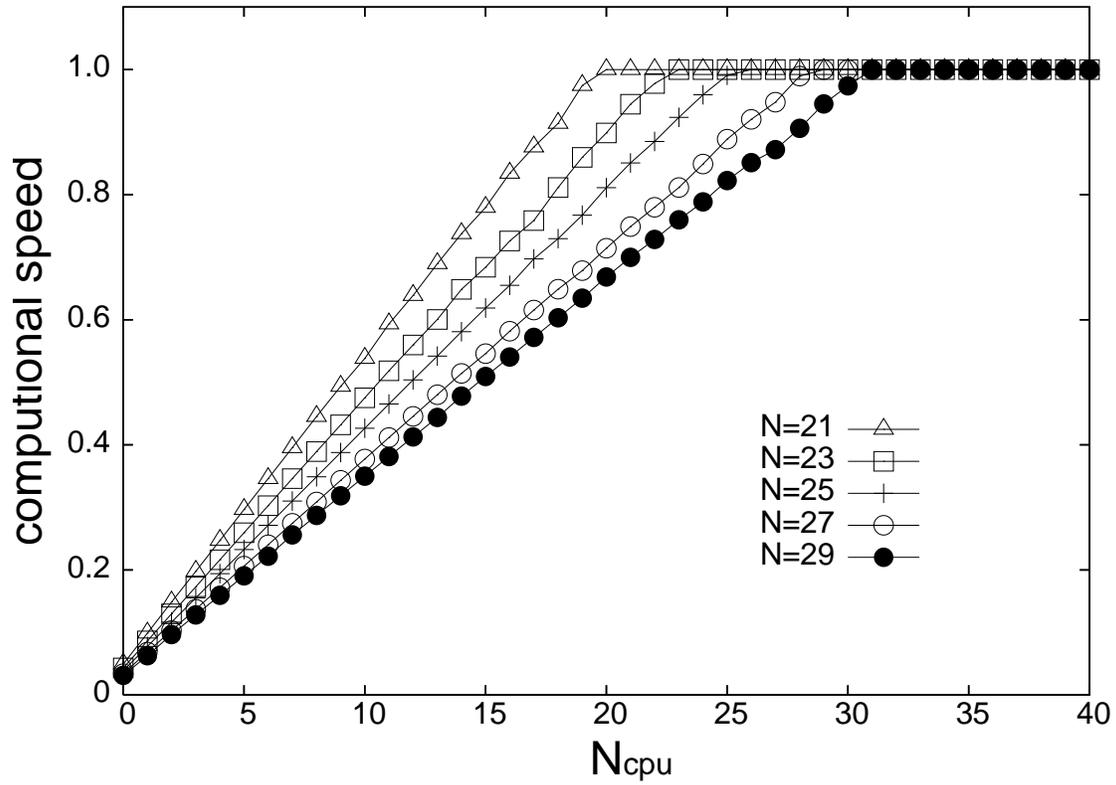}
\caption{The computational speed as the function of the number of computational nodes.
The computational speed is defined as the inverse computational time,
normalized so that its saturated value is 1.00.}
\label{scale}
\end{figure}

By the efficient distribution of the enumeration tasks among the computational nodes, the computational speed scales linearly with the number of computational nodes until the saturation occurs. It is obvious that the number of nodes used, $N_{\rm CPU}$, cannot exceed the number of boxes, but the saturation occurs at a smaller value of $N_{\rm CPU}$ because the enumeration time varies widely among the boxes. A few nodes enumerating the box with large number of conformations tend to keep computing even after most of the  nodes have completed computations, causing the deviation from the linear scaling of computation speed with $N_{\rm CPU}$.
By recording the computation times for each boxes and assuming the most efficient distribution of tasks between the nodes, the computation time could be calculated as a function of $N_{\rm CPU}$. The speed of computation, the inverse of the computation time, is plotted in Figure~\ref{scale} for various values of $N$, where the saturated value of the speed is normalized to one.
We see that the linear scaling holds for up to  $N_{\rm CPU} = N_{\rm CPU}^{({\rm max})}$ which increases with $N$.

\begin{figure}
\includegraphics*[width=.9\textwidth]{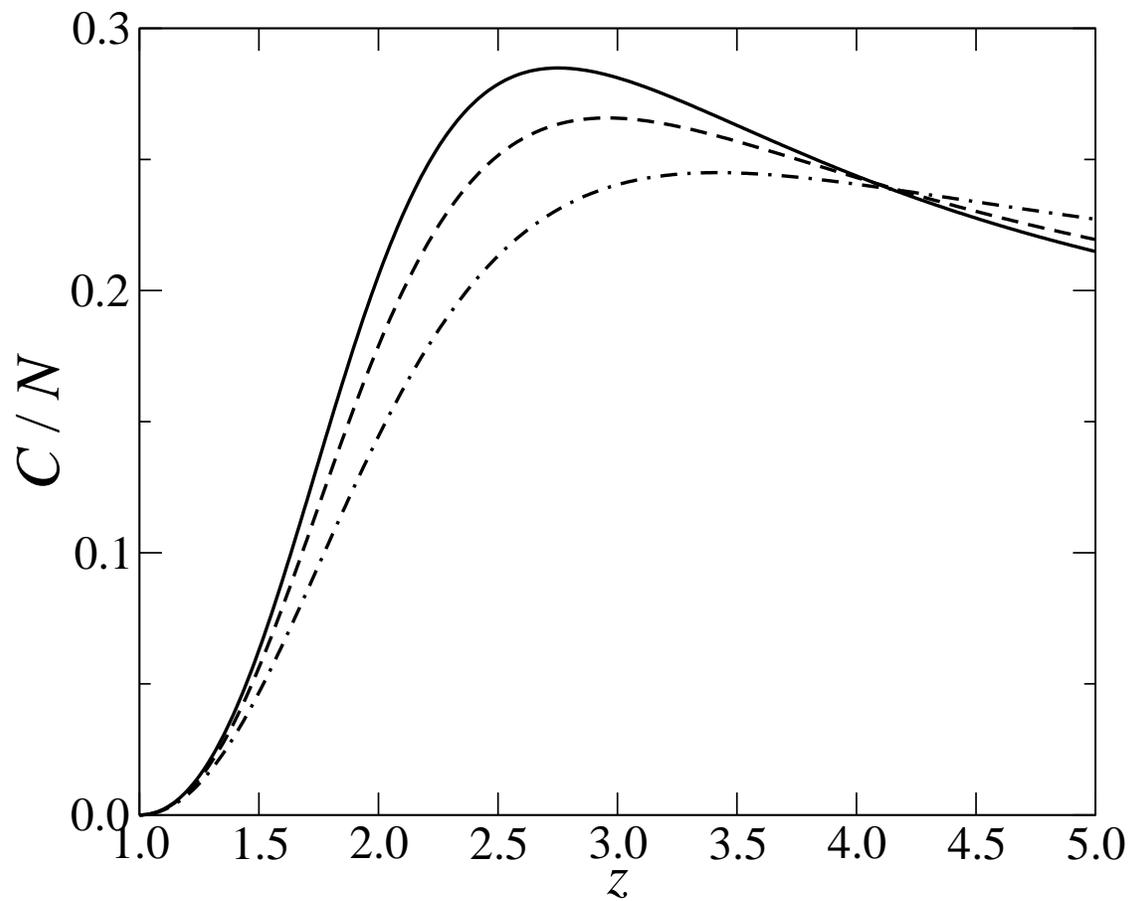}
\caption{The specific heat for $N = 20$, 28, and 36 from bottom to top.}
\label{sh}
\end{figure}

\begin{figure}
\includegraphics*[width=.9\textwidth]{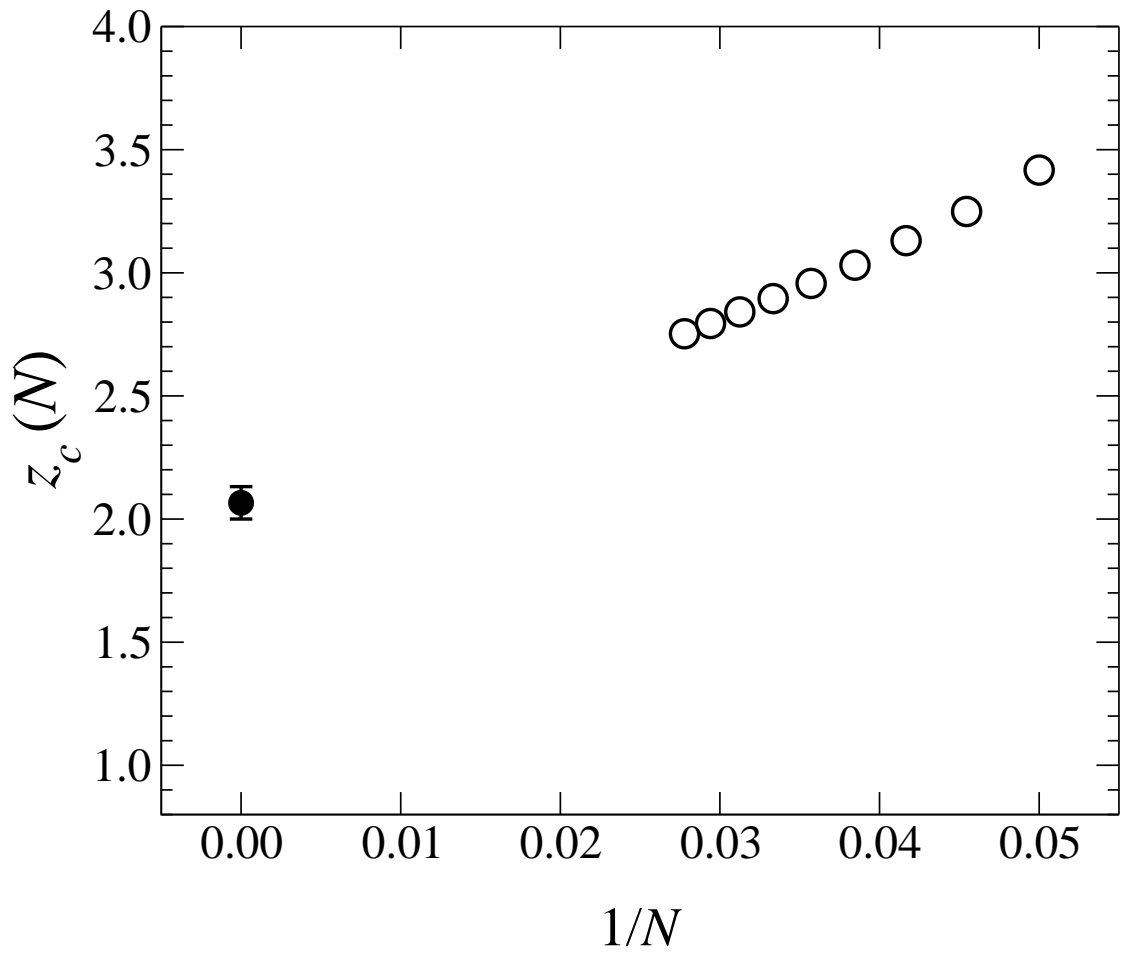}
\caption{The finite value approximation of the transition point $z_c$
and its extrapolated value at $N = \infty$.}
\label{sh_bst}
\end{figure}

As an example of the application of our method, we calculate the specific heat per monomer,
\be
C(T,N)/\epsilon^2 N = \frac{1}{\epsilon^2 N}\frac{\partial E}{\partial T} = \frac{\beta^2}{\epsilon^2 N} \frac{\partial^2 \ln Z}{\partial \beta^2} = \frac{(\ln z)^2}{N}\left[ \frac{\sum_k k^2 \Omega(k) z^k}{\sum_p \Omega(p) z^p}-\left(\frac{\sum_k k \Omega(k) z^k}{\sum_p \Omega(p) z^p}\right)^2\right]
\ee
which is plotted in Figure~\ref{sh} as a function of $z$ for several values of $N$.
The finite $N$ approximation of the transition point, $z_c(N)$, is obtained from the condition
$\frac{\partial C}{\partial z} = 0$. We observe a peak around $z \simeq 2$, which becomes sharper as $N$ increases.
The point of the collapse transition is known to follow the finite-size scaling
\begin{equation}
    z_c(N) - z_c(\infty) \sim N^{-\phi},
\label{zc}
\end{equation}
where $\phi$ is the crossover exponent whose exact value is believed to be $3/7$~\cite{DS87}.
We apply the Bulirsch-Stoer extrapolation~\cite{BS64,PTVF92,M02} to Eq.~(\ref{zc})
and then obtain $z_c(\infty) = 2.07(7)$, equivalent to $T_c/\epsilon = 1.37(7)$,
where the data for even $N$ with $20 \le  N \le 36$ were used.
$z_c(N)$ is displayed in Figure~\ref{sh_bst} as the functions of $1/N$,
along with the extrapolated value $z_c(\infty)$.
The current result is in reasonable agreement with those from the earlier works.


\section{Discussion}

We developed an efficient parallel algorithm for calculating the exact partition function of a lattice polymer.
An efficient parallelization of
the calculation was achieved by classifying the  conformations
according to the shape of the box spanned by a conformation. Only the conformations corresponding to a given box were enumerated at a time, pruning partial conformations incompatible with the box at an early stage. The calculation time was further reduced by preventing the conformations
related by symmetries from being generated more than once.
The parallel efficiency could be maximized by requiring that  any node that finishes the task of enumeration for a box takes over a new task of enumeration for another box whose conformations have not been enumerated by any of the nodes.
As an illustration of applications, we studied the collapse transition of lattice homopolymers in square lattices,
by calculating the specific heat. The exact partition function can also be used for calculating the partition function zeros in the complex temperature plane, which is a more sensitive indicator of the phase transition than the specific heat~\cite{LKL10}.

As mentioned in the previous section, the linear scaling of the computation speed with the number of CPUs $N_{\rm CPU}$ breaks down well before $N_{\rm CPU}$ reaches the total number of boxes, due to the fact the computation time varies wildly depending on boxes. However, considering the fact that $N_{\rm CPU}$ cannot be made arbitrarily large in practice, the saturation effect can be neglected in many situations, particularly when long chains are studied, since  the saturation limit $N_{\rm CPU}^{({\rm max})}$ grows with the chain length $N$ (See the Result). As can be seen from the Result (See Fig.5), the partition function of the two-dimensional polymer chain of length 29 can be enumerated with up to 30 CPUs without saturation of the linear scaling, and one can employ more CPUs for studying chains of longer lengths. Since linear scaling becomes better as the chain length increases,  our method is a powerful tool for studying long polymer chains with limited computational resources.

An idea of classifying conformations of two-dimensional homopolymers according to spanning boxes,
similar to ours in certain aspects, has been introduced for the calculation of the partition function of a polymer on the square lattice at infinite temperature using transfer matrix formalism~\cite{J04}.
At least for such a calculation the transfer matrix was claimed to be superior to the direct counting,
and it would be interesting to see whether it can be generalized for calculating the partition function at arbitrary temperatures in two dimensions, without introducing much additional computational costs.
The current algorithm is more general in that not only the partition function at arbitrary temperature can be calculated, but heteropolymers and arbitrary dimensions can be treated in a straightforward manner.
The explicit applications of the algorithm to these cases are left for future works.

\begin{acknowledgments}
This work was supported by Mid-career Researcher Program through NRF grant funded by the MEST (No.2010-0000220).
\end{acknowledgments}


\newpage

\begin{table}
\caption{The number of conformations $\omega(K)$ on a square lattice as a function of the chain length $29 \le N \le 32$ and the number of contacts $K$.}
\begin{tabular}{c|r|r|r|r|}
\hline
\hline
\backslashbox{$K$}{$N$}
    &29~~~~~~~~~    &30~~~~~~~~~    &31~~~~~~~~~    &32~~~~~~~~~    \\
\hline
~~0 &7689321701     &17982126658    &42108189098    &98421806691    \\
~~1 &23540565448    &56977682194    &137862646874   &332762640146   \\
~~2 &40085909835    &100142920787   &249700259569   &620847470396   \\
~~3 &49071555164    &126359874347   &324279253784   &829487165382   \\
~~4 &48771398860    &129152420800   &340420363856   &894055572891   \\
~~5 &41585979484    &113114869570   &305828397226   &823666234004   \\
~~6 &31508871807    &87958369563    &243684809928   &672408321619   \\
~~7 &21617211324    &61936948230    &175770392578   &496787988274   \\
~~8 &13642191086    &40119932593    &116592999110   &337522862616   \\
~~9 &7980014626     &24122755170    &71817661842    &213018321049   \\
10  &4362872816     &13562820674    &41372702687    &125886832996   \\
\hline
11  &2226147024     &7152132929     &22391023650    &69957624306    \\
12  &1065063681     &3542639525     &11397385748    &36690896460    \\
13  &475157642      &1646914139     &5456762684     &18158394435    \\
14  &194929001      &716011778      &2455124926     &8476869526     \\
15  &72870960       &286636733      &1024918738     &3722404274     \\
16  &24595083       &104686477      &393687071      &1519486734     \\
17  &6751332        &34676719       &136943328      &571834936      \\
18  &899613         &9090306        &41715633       &194513054      \\
19  &16294          &1005977        &8866818        &57891860       \\
20  &               &13498          &636771         &11290845       \\
21  &               &               &               &656376         \\
\hline
Total & 293922322781 & 784924528667 & 2092744741919 &  5584227078870 \\
\hline
\hline
\end{tabular}
\label{29to32}
\end{table}
\begin{table}
\caption{The number of conformations $\omega(K)$ on a square lattice as a function of the chain length $33 \le N \le 36$ and the number of contacts $K$.}
\begin{tabular}{c|r|r|r|r|}
\hline
\hline
\backslashbox{$K$}{$N$}
    &33~~~~~~~~~    &34~~~~~~~~~    &35~~~~~~~~~    &36~~~~~~~~~    \\
\hline
~~0 &230322480773   &538091763166   &1258493243324  &2938908879305  \\
~~1 &802996241232   &1933501499531  &4654740470620  &11183769131112 \\
~~2 &1541179904025  &3815982548374  &9435546145403  &23276263902178 \\
~~3 &2115681496986  &5380293120526  &13648570547310 &34530182956163 \\
~~4 &2338726472743  &6097828689993  &15844205741060 &41046002582789 \\
~~5 &2206962182754  &5892600481193  &15663662724418 &41502969876897 \\
~~6 &1843564131938  &5035936926127  &13681123644389 &37041494009397 \\
~~7 &1393031891232  &3891460627050  &10798327608646 &29859153057072 \\
~~8 &967612822630   &2763594584708  &7829412575160  &22103175375035 \\
~~9 &624333988838   &1823442056984  &5273396514658  &15198342307484 \\
10  &377300221611   &1127227716216  &3327749067600  &9793121298664  \\
\hline
11  &214525075976   &656411900406   &1978790499320  &5948956055099  \\
12  &115237903606   &361505461087   &1113409518803  &3422967752173  \\
13  &58512766632    &188687694809   &594317169744   &1870659345753  \\
14  &28078596777    &93399109108    &301280956584   &972915151671   \\
15  &12713385376    &43768480489    &144908527014   &481649031145   \\
16  &5394346351     &19397612212    &66030540012    &226685149424   \\
17  &2118111650     &8050408136     &28387608698    &101251448207   \\
18  &763529938      &3106194291     &11380453744    &42658656267    \\
19  &244420464      &1095275398     &4222218392     &16800266331    \\
20  &64896504       &342819455      &1417505104     &6127509709     \\
\hline
21  &9564594        &87997218       &412844504      &2004417664     \\
22  &306498         &11551406       &89824129       &573845730      \\
23  &               &265502         &8277188        &118501239      \\
24  &               &               &105265         &9156136        \\
25  &               &               &               &57337          \\
\hline
Total &  14879374739128 &  39675824783385 & 105659884331089 &  281566759719981 \\
\hline
\hline
\end{tabular}
\label{33to36}
\end{table}
\newpage

\end{document}